\documentclass[twocolumn,english,prl,showkeys]{revtex4-1}

\usepackage[T1]{fontenc}
\usepackage[latin9]{inputenc}
\usepackage{amsmath}
\usepackage{graphicx}

\makeatletter
\usepackage{babel}

\usepackage{babel}

\makeatother

\usepackage{babel}
\begin{document}

\title{Solving the Quantum Many-Body Problem with Artificial Neural Networks}

\author{Giuseppe Carleo}

\email{gcarleo@ethz.ch}

\selectlanguage{english}%

\affiliation{Theoretical Physics, ETH Zurich, 8093 Zurich, Switzerland }

\author{Matthias Troyer}

\affiliation{Theoretical Physics, ETH Zurich, 8093 Zurich, Switzerland }

\affiliation{Quantum Architectures and Computation Group, Microsoft Research,
Redmond, WA 98052, USA}

\affiliation{Station Q, Microsoft Research, Santa Barbara, CA 93106-6105, USA}
\begin{abstract}
\textbf{The challenge posed by the many-body problem in quantum physics
originates from the difficulty of describing the non-trivial correlations
encoded in the exponential complexity of the many-body wave function.
Here we demonstrate that systematic machine learning of the wave function
can reduce this complexity to a tractable computational form, for
some notable cases of physical interest. We introduce a variational
representation of quantum states based on artificial neural networks
with variable number of hidden neurons. A reinforcement-learning scheme
is then demonstrated, capable of either finding the ground-state or
describing the unitary time evolution of complex interacting quantum
systems. We show that this approach achieves very high accuracy in
the description of equilibrium and dynamical properties of prototypical
interacting spins models in both one and two dimensions, thus offering
a new powerful tool to solve the quantum many-body problem. }
\end{abstract}
\maketitle
The wave function $\Psi$ is the fundamental object in quantum physics
and possibly the hardest to grasp in a classical world. $\Psi$ is
a monolithic mathematical quantity that contains all the information
on a quantum state, be it a single particle or a complex molecule.
In principle, an exponential amount of information is needed to fully
encode a generic many-body quantum state. However, Nature often proves
herself benevolent, and a wave function representing a \emph{physical}
many-body system can be typically characterized by an amount of information
much smaller than the maximum capacity of the corresponding Hilbert
space. A limited amount of quantum entanglement, as well as the typicality
of a small number of physical states, are then the blocks on which
modern approaches build upon to solve the many-body Schr\"{o}dinger's
equation with a limited amount of classical resources.

Numerical approaches directly relying on the wave function can either
sample a finite number of physically relevant configurations or perform
an efficient \emph{compression} of the quantum state. Stochastic approaches,
like quantum Monte Carlo (QMC) methods, belong to the first category
and rely on probabilistic frameworks typically demanding a positive-semidefinite
wave function. \cite{ceperley1986quantum,foulkes2001quantum,carlson2015quantum}.\textbf{
}Compression approaches instead rely on efficient representations
of the wave function, and most notably in terms of matrix product
states (MPS) \cite{white1992density,rommer1997classof,schollwock2011thedensitymatrix}
or more general tensor networks \cite{orus2014apractical,verstraete2008matrixproduct}.
Examples of systems where existing approaches fail are however numerous,
mostly due to the sign problem in QMC \cite{troyer2005computational},
and to the inefficiency of current compression approaches in high-dimensional
systems. As a result, despite the striking success of these methods,
a large number of unexplored regimes exist, including many interesting
open problems. These encompass fundamental questions ranging from
the dynamical properties of high-dimensional systems \cite{polkovnikov2011colloquium,j.eisert2015quantum}
to the exact ground-state properties of strongly interacting fermions
\cite{montorsi1992thehubbard,thouless1972thequantum}. At the heart
of this lack of understanding lyes the difficulty in finding a general
strategy to reduce the exponential complexity of the full many-body
wave function down to its most essential features \cite{freericks2014thenonequilibrium}.

In a much broader context, the problem resides in the realm of dimensional
reduction and feature extraction. Among the most successful techniques
to attack these problems, artificial neural networks play a prominent
role \cite{hinton2006reducing}. They can perform exceedingly well
in a variety of contexts ranging from image and speech recognition
\cite{lecun2015deeplearning} to game playing \cite{silver2016mastering}.
Very recently, applications of neural network to the study of physical
phenomena have been introduced \cite{schoenholz2016astructural,carrasquilla2016machine,wang2016discovering}.
These have so-far focused on the classification of complex phases
of matter, when exact sampling of configurations from these phases
is possible. The challenging goal of solving a many-body problem without
prior knowledge of exact samples is nonetheless still unexplored and
the potential benefits of Artificial Intelligences in this task are
at present substantially unknown. It appears therefore of fundamental
and practical interest to understand whether an artificial neural
network can modify and adapt itself to describe and analyze a quantum
system. This ability could then be used to solve the quantum many-body
problem in those regimes so-far inaccessible by existing exact numerical
approaches.

Here we introduce a representation of the wave function in terms of
artificial neural networks specified by a set of internal parameters
$\mathcal{W}$. We present a stochastic framework for reinforcement
learning of the parameters $\mathcal{W}$ allowing for the best possible
representation of both ground-state and time-dependent physical states
of a given quantum Hamiltonian $\mathcal{H}$. The parameters of the
neural network are then optimized (trained, in the language of neural
networks) either by static variational Monte Carlo (VMC) sampling
\cite{mcmillan1965groundstate}, or in time-dependent VMC \cite{carleo2012localization,carleo2014lightcone},
when dynamical properties are of interest. We validate the accuracy
of this approach studying the Ising and Heisenberg models in both
one and two-dimensions. The power of the \emph{neural-network quantum
states} (NQS) is demonstrated obtaining state-of-the-art accuracy
in both ground-state and out-of-equilibrium dynamics. In the latter
case, our approach effectively solves the phase-problem traditionally
affecting stochastic Quantum Monte Carlo approaches, since their introduction.
\begin{figure}
\includegraphics[width=0.85\columnwidth]{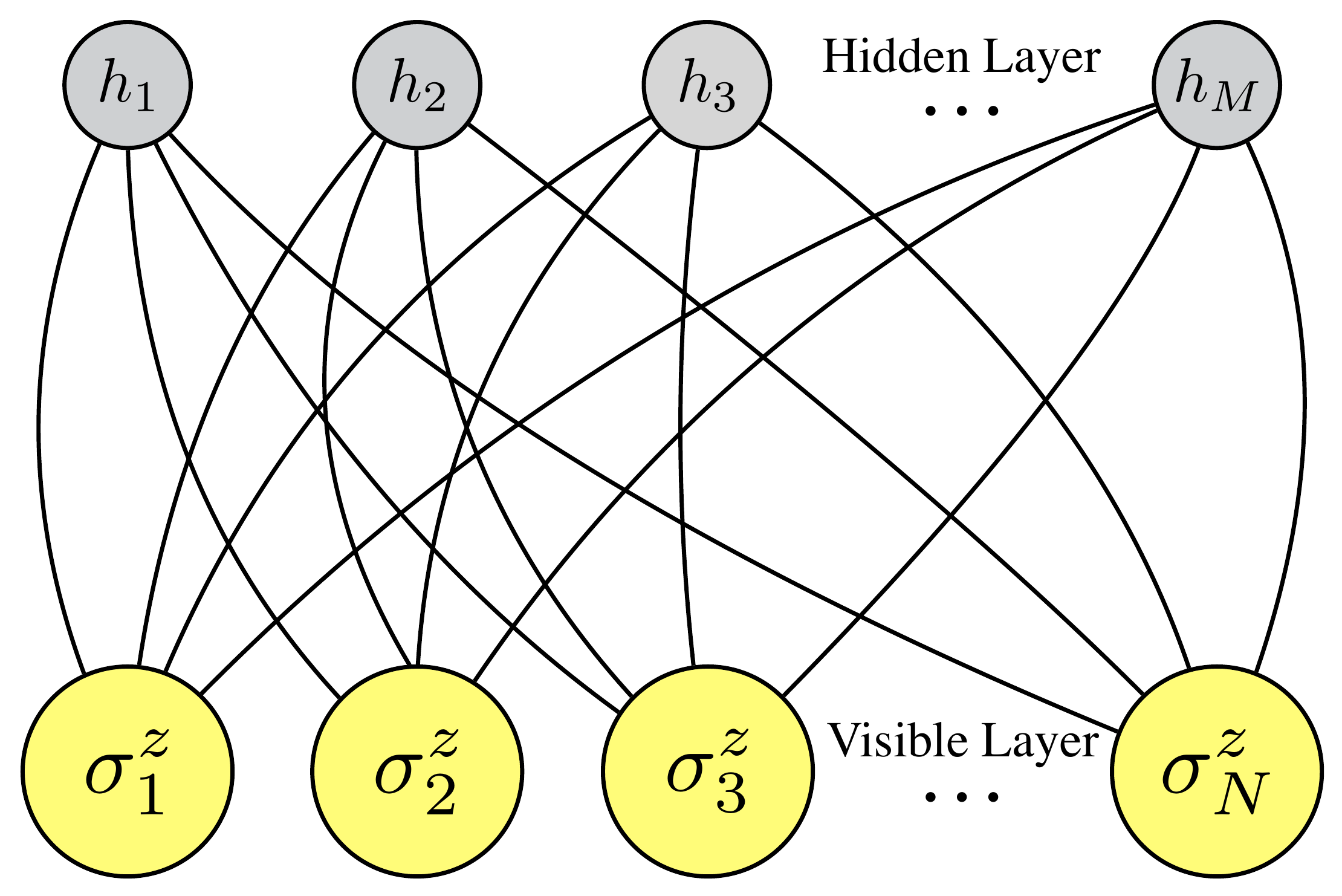}

\caption{\label{fig:Artificial-Neural-network}\textbf{Artificial Neural network
encoding a many-body quantum state of $N$ spins.} Shown is a restricted
Boltzmann machine architecture which features a set of $N$ visible
artificial neurons (yellow dots) and a set of $M$ hidden neurons
(grey dots). For each value of the many-body spin configuration $\mathcal{S}=(\sigma_{1}^{z},\sigma_{2}^{z},\dots\sigma_{N}^{z})$,
the artificial neural network computes the value of the wave function
$\Psi(\mathcal{S})$.}
\end{figure}

\begin{figure*}
\includegraphics[clip,height=6cm]{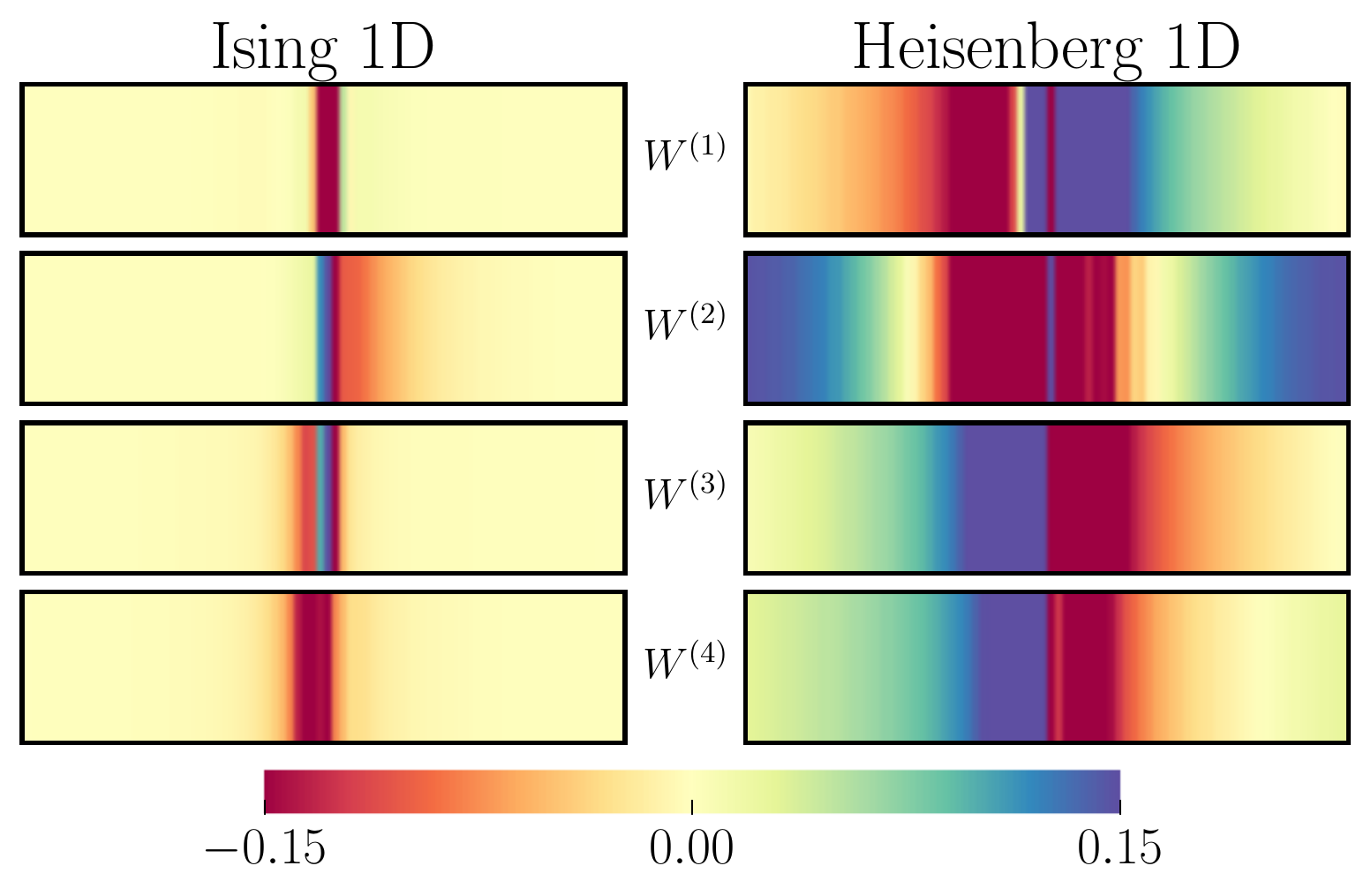}\hspace{2cm}\includegraphics[clip,height=6cm]{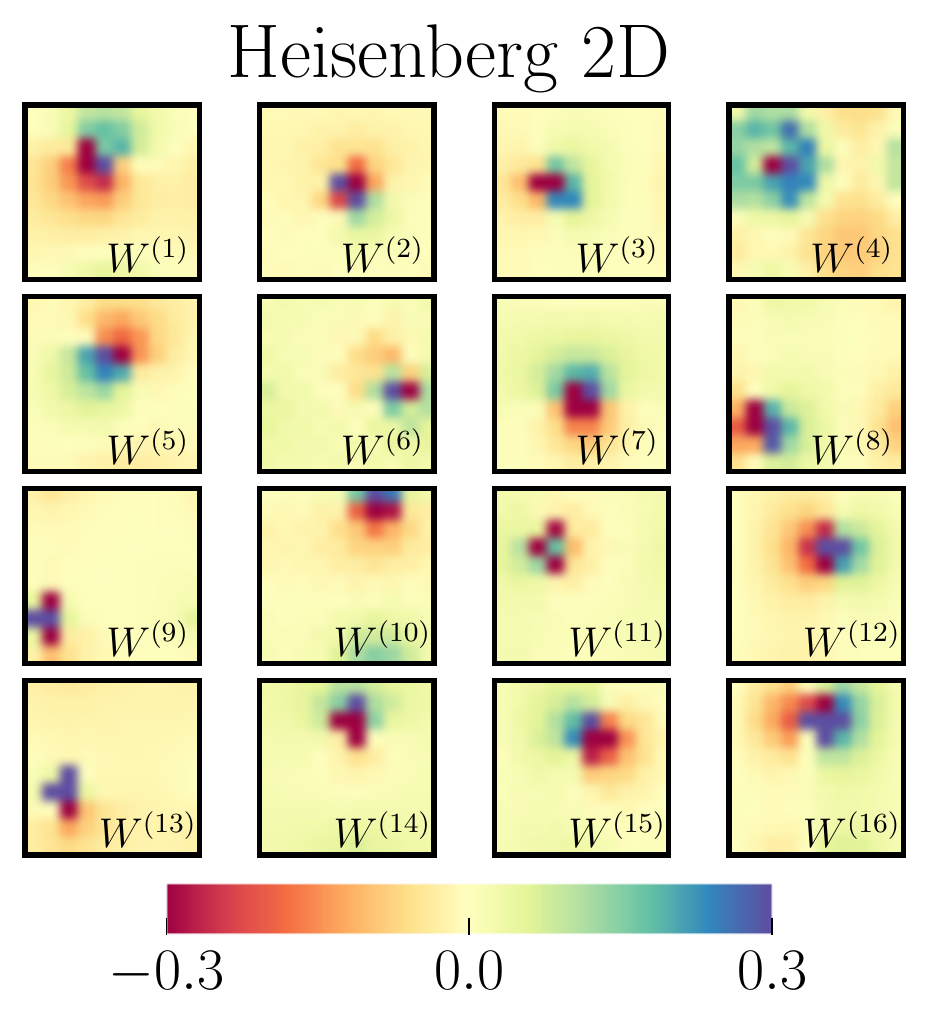}

\caption{\label{fig:GS-networks} \textbf{Neural Network representation of
the many-body ground states of prototypical spin models in one and
two dimensions.} In the left group of panels we show the feature maps
for the one-dimensional TFI model at the critical point $h=1$, as
well as for the AFH model. In both cases the hidden-unit density is
$\alpha=4$ and the lattices comprise $80$ sites. Each horizontal
colormap shows the values that the $f$-th feature map $W_{j}^{(f)}$
takes on the $j$-th lattice site (horizontal axis, broadened along
the vertical direction for clarity). In the right group of panels
we show the feature maps for the two-dimensional Heisenberg model
on a square lattice, for $\alpha=16$. In this case the the horizontal
(vertical) axis of the colormaps correspond to the $x$($y$) coordinates
on a $10\times10$ square lattice. Each of the feature maps act as
effective filters on the spin configurations, capturing the most important
quantum correlations.}
\end{figure*}

\paragraph*{Neural-Network Quantum States ---}

Consider a quantum system with $N$ discrete-valued degrees of freedom
$\mathcal{S}=(\mathcal{S}_{1},\mathcal{S}_{2}\dots\mathcal{S}_{N})$,
which may be spins, bosonic occupation numbers, or similar. The many-body
wave function is a mapping of the $N-$dimensional set $\mathcal{S}$
to (exponentially many) complex numbers which fully specify the amplitude
and the phase of the quantum state. The point of view we take here
is to interpret the wave function as a computational black box which,
given an input many-body configuration $\mathcal{S}$, returns a phase
and an amplitude according to $\Psi(\mathcal{S})$. Our goal is to
approximate this computational black box with a neural network, trained
to best represent $\Psi(\mathcal{S})$. Different possible choices
for the artificial neural-network architectures have been proposed
to solve specific tasks, and the best architecture to describe a many-body
quantum system may vary from one case to another. For the sake of
concreteness, in the following we specialize our discussion to restricted
Boltzmann machines (RBM) architectures, and apply them to describe
spin $1/2$ quantum systems. In this case, RBM artificial networks
are constituted by one visible layer of $N$ nodes, corresponding
to the physical spin variables in a chosen basis (say for example
$\mathcal{S}=\sigma_{1}^{z},\dots\sigma_{N}^{z}$) , and a single
hidden layer of $M$ auxiliary spin variables ($h_{1}\dots h_{M}$)
(see Fig. \ref{fig:Artificial-Neural-network}). This description
corresponds to a variational expression for the quantum states which
reads:
\[
\Psi_{M}(\mathcal{S};\mathcal{W})=\sum_{\{h_{i}\}}e^{\sum_{j}a_{j}\sigma_{j}^{z}+\sum_{i}b_{i}h_{i}+\sum_{ij}W_{ij}h_{i}\sigma_{j}^{z}},
\]
where $h_{i}=\{-1,1\}$ is a set of $M$ hidden spin variables, and
the weights $\mathcal{W}=\{a_{i},b_{j},W_{ij}\}$ fully specify the
response of the network to a given input state $\mathcal{S}$. Since
this architecture features no intra-layer interactions, the hidden
variables can be explicitly traced out, and the wave function reads
$\Psi(\mathcal{S};\mathcal{W})=e^{\sum_{i}a_{i}\sigma_{i}^{z}}\times\Pi_{i=1}^{M}F_{i}(\mathcal{S}),$
where $F_{i}(\mathcal{S})=2\cosh\left[b_{i}+\sum_{j}W_{ij}\sigma_{j}^{z}\right]$.
The network weights are, in general, to be taken complex-valued in
order to provide a complete description of both the amplitude and
the wave-function's phase.

The mathematical foundations for the ability of NQS to describe intricate
many-body wave functions are the numerously established representability
theorems \cite{kolmogorov1961onthe,hornik1991approximation,leroux2008representational},
which guarantee the existence of network approximates of high-dimensional
functions, provided a sufficient level of smoothness and regularity
is met in the function to be approximated. Since in most physically
relevant situations the many-body wave function reasonably satisfies
these requirements, we can expect the NQS form to be of broad applicability.
One of the practical advantages of this representation is that its
quality can, in principle, be systematically improved upon increasing
the number of hidden variables. The number $M$ (or equivalently the
density $\alpha=M/N$) then plays a role analogous to the bond dimension
for the MPS. Notice however that the correlations induced by the hidden
units are intrinsically non local in space and are therefore well
suited to describe quantum systems in arbitrary dimension. Another
convenient point of the NQS representation is that it can be formulated
in a symmetry-conserving fashion. For example, lattice translation
symmetry can be used to reduce the number of variational parameters
of the NQS ansatz, in the same spirit of shift-invariant RBM's \cite{sohn2012learning,norouzi2009stacksof}.
Specifically, for integer hidden variable density $\alpha=1,2,\dots$,
the weight matrix takes the form of feature \emph{filters} $W_{j}^{(f)}$
, for $f\in[1,\alpha]$. These filters have a total of $\alpha N$
variational elements \emph{in lieu} of the $\alpha N^{2}$ elements
of the asymmetric case (see Supp. Mat. for further details).

Given a general expression for the quantum many-body state, we are
now left with the task of solving the many-body problem upon machine
learning of the network parameters $\mathcal{W}$. In the most interesting
applications the exact many-body state is unknown, and it is typically
found upon solution either of the static Schr\"{o}dinger equation $\mathcal{H}\left|\Psi\right\rangle =E\left|\Psi\right\rangle $,
either of the time-dependent one $i\mathcal{H}\left|\Psi\right(t)\rangle=\frac{d}{dt}\left|\Psi(t)\right\rangle $,
for a given Hamiltonian $\mathcal{H}$. In the absence of samples
drawn according to the exact wave function, supervised learning of
$\Psi$ is therefore not a viable option. Instead, in the following
we derive a consistent reinforcement learning approach, in which either
the ground-state wave function or the time-dependent one are learned
on the basis of feedback from variational principles.

\paragraph*{Ground State ---}

To demonstrate the accuracy of the NQS in the description of complex
many-body quantum states, we first focus on the goal of finding the
best neural-network representation of the unknown ground state of
a given Hamiltonian $\mathcal{H}$. In this context, reinforcement
learning is realized through minimization of the expectation value
of the energy $E(\mathcal{W})=\langle\Psi_{M}|\mathcal{H}|\Psi_{M}\rangle/\langle\Psi_{M}|\Psi_{M}\rangle$
with respect to the network weights $\mathcal{W}$. In the stochastic
setting, this is achieved with an iterative scheme. At each iteration
$k$, a Monte Carlo sampling of $\left|\Psi_{M}(S;\mathcal{W}_{k})\right|^{2}$
is realized, for a given set of parameters $\mathcal{W}_{k}$. At
the same time, stochastic estimates of the energy gradient are obtained.
These are then used to propose a next set of weights $\mathcal{W}_{k+1}$
with an improved gradient-descent optimization \cite{sorella2007weakbinding}.
The overall computational cost of this approach is comparable to that
of standard ground-state Quantum Monte Carlo simulations (see Supp.
Material).
\begin{figure*}[t]
\includegraphics[clip,width=0.8\paperwidth]{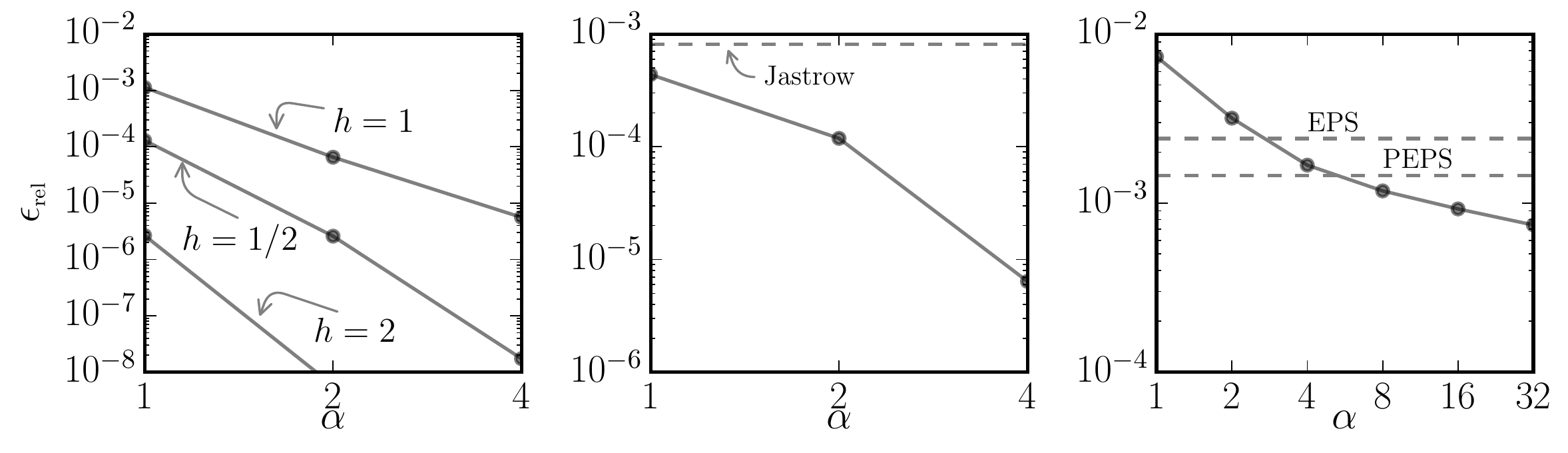}

\caption{\label{fig:Gs-Accuracy}\textbf{Finding the many-body ground-state
energy with neural-network quantum states.} Shown is the error of
the NQS ground-state energy relative to the exact value, for several
test cases. Arbitrary precision on the ground-state energy can be
obtained upon increasing the hidden units density, $\alpha$. (Left
panel) Accuracy for the one-dimensional TFI model, at a few values
of the field strength $h$, and for a $80$ spins chain with PBC.
Points below $10^{-8}$ are not shown to easy readability. (Central
panel) Accuracy for the one-dimensional AFH model, for a $80$ spins
chain with PBC, compared to the Jastrow ansatz (horizontal dashed
line). (Right panel) Accuracy for the AFH model on a $10\times10$
square lattice with PBC, compared to the precision obtained by EPS
(upper dashed line) and PEPS (lower dashed line). For all cases considered
here the NQS description reaches MPS-grade accuracies in 1D, while
it systematically improves the best known variational states for 2D
finite lattice systems. }
\end{figure*}

To validate our scheme, we consider the problem of finding the ground
state of two prototypical spin models, the transverse-field Ising
(TFI) model and the anti-ferromagnetic Heisenberg (AFH) model. Their
Hamiltonians are
\begin{equation}
\mathcal{H}_{\text{TFI}}=-h\sum_{i}\sigma_{i}^{x}-\sum_{\langle i,j\rangle}\sigma_{i}^{z}\sigma_{j}^{z}
\end{equation}
and
\begin{equation}
\mathcal{H}_{\text{AFH}}=\sum_{\left\langle i,j\right\rangle }\sigma_{i}^{x}\sigma_{j}^{x}+\sigma_{i}^{y}\sigma_{j}^{y}+\sigma_{i}^{z}\sigma_{j}^{z},
\end{equation}
respectively, where $\sigma^{x},\sigma^{y},\sigma^{z}$ are Pauli
matrices.

In the following, we consider the case of both one and two dimensional
lattices with periodic boundary conditions (PBC). In Fig. \ref{fig:GS-networks}
we show the optimal network structure of the ground states of the
two spin models for a hidden variables density $\alpha=4$ and with
imposed translational symmetries. We find that each filter $f=[1,\dots\alpha]$
learns specific correlation features emerging in the ground state
wave function. For example, in the 2D case it can be seen (Fig. \ref{fig:GS-networks},
rightmost panels) how the neural network learns patterns corresponding
to anti-ferromagnetic correlations. The general behavior of the NQS
is completely analogous to what observed in convolutional neural networks,
where different layers learn specific structures of the input data.

In Fig. \ref{fig:Gs-Accuracy} we show the accuracy of the NQS states,
quantified by the relative error on the ground-state energy $\epsilon_{\textrm{rel}}=\left(E_{\textrm{NQS}}(\alpha)-E_{\textrm{exact}}\right)/\left|E_{\textrm{exact }}\right|$,
for several values of $\alpha$ and model parameters. In the left
panel, we compare the variational NQS energies with the exact result
obtained by fermionization of the TFI model, on a one-dimensional
chain with PBC. The most striking result is that NQS achieve a controllable
and arbitrary accuracy which is compatible with a power-law behavior
in $\alpha$. The hardest to learn ground-state is at the quantum
critical point $h=1$, where nonetheless a remarkable accuracy of
one part per million can be easily achieved with a relatively modest
density of hidden units. The same remarkable accuracy is obtained
for the more complex one-dimensional AFH model (central panel). In
this case we observe as well a systematic drop in the ground-state
energy error, which for a small $\alpha=4$ attains the same very
high precision obtained for the TFI model at the critical point. Our
results are compared with the accuracy obtained with the spin-Jastrow
ansatz (dashed line in the central panel), which we improve by several
orders of magnitude. It is also interesting to compare the value of
$\alpha$ with the MPS bond dimension $M$, needed to reach the same
level of accuracy. For example, on the AFH model with PBC, we find
that with a standard DMRG implementation \cite{dolfi2014matrixproduct}
we need $M\sim160$ to reach the accuracy we have at $\alpha=4$.
This points towards a more compact representation of the many-body
state in the NQS case, which features about $3$ orders of magnitude
less variational parameters than the corresponding MPS ansatz.

We next study the AFH model on a two-dimensional square lattice, comparing
in the right panel of Fig. \ref{fig:Gs-Accuracy} to QMC results \cite{sandvik1997finitesize}.
As expected from entanglement considerations, the 2D case proves harder
for the NQS. Nonetheless, we always find a systematic improvement
of the variational energy upon increasing $\alpha$, qualitatively
similar to the 1D case. The increased difficulty of the problem is
reflected in a slower convergence. We still obtain results at the
level of existing state-of-the-art methods or better. In particular,
with a relatively small hidden unit density $(\alpha\sim4)$ we already
obtain results at the same level than the best known variational ansatz
to-date for finite clusters (the EPS of Ref. \cite{mezzacapo2009groundstate}
and the PEPS states of Ref. \cite{lubasch2014algorithms}). Further
increasing $\alpha$ then leads to a sizable improvement and consequently
yields the best variational results so-far-reported for this 2D model
on finite lattices.

\paragraph*{Unitary Dynamics ---}

NQS are not limited to ground-state problems but can be extended to
the time-dependent Schr\"{o}dinger equation. For this purpose we define
complex-valued and time-dependent network weights $\mathcal{W}(t)$
which at each time $t$ are trained to best reproduce the quantum
dynamics, in the sense of the Dirac-Frenkel time-dependent variational
principle \cite{dirac1930noteon,frenkel1934wavemechanics}. In this
context, the variational residuals
\begin{equation}
R(t;\dot{\mathcal{W}}(t))=\mathrm{dist}(\partial_{t}\Psi(\mathcal{W}(t)),-i\mathcal{H}\Psi)
\end{equation}
are the objective functions to be minimized as a function of the time
derivatives of the weights $\dot{\mathcal{W}}(t)$ (see Supp. Mat.)
In the stochastic framework, this is achieved by a time-dependent
VMC method \cite{carleo2012localization,carleo2014lightcone}, which
samples $\left|\Psi_{M}(S;\mathcal{W}(t))\right|^{2}$ at each time
and provides the best stochastic estimate of the $\dot{\mathcal{W}}(t)$
that minimize $R^{2}(t)$, with a computational cost $\mathcal{O}(\alpha N^{2})$.
Once the time derivatives determined, these can be conveniently used
to obtain the full time evolution after time-integration.
\begin{figure*}[t]
\includegraphics[clip,width=1.6\columnwidth]{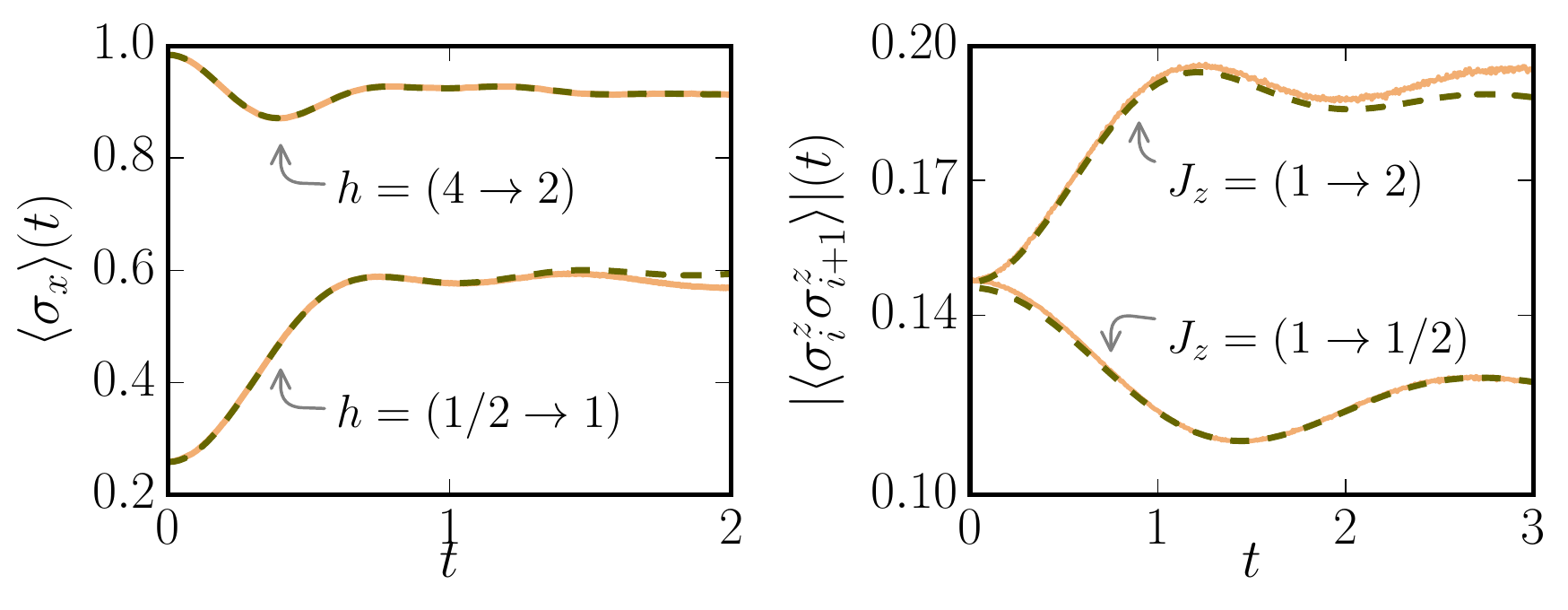}

\caption{\label{fig:Time-Accuracy}\textbf{Describing the many-body unitary
time evolution with neural-network quantum states.} Shown are results
for the time evolution induced by a quantum quench in the microscopical
parameters of the models we study (the transverse field $h$, for
the TFI model and the coupling constant $J_{z}$ in the AFH model).
(Left Panel) NQS results (solid lines) are compared to exact results
for the transverse spin polarization in the one-dimensional TFI model
(dashed lines). (Right Panel) In the AFH model, the time-dependent
nearest-neighbors spin correlations are compared to exact numerical
results obtained with t-DMRG for an open one-dimensional chain representative
of the thermodynamic limit (dashed lines).}
\end{figure*}

To demonstrate the effectiveness of the NQS in the dynamical context,
we consider the unitary dynamics induced by quantum quenches in the
coupling constants of our spin models. In the TFI model we induce
a non-trivial quantum dynamics by means of an instantaneous change
in the transverse field: the system is initially prepared in the ground-state
of the TFI model for some transverse field, $h_{i}$, and then let
evolve under the action of the TFI Hamiltonian with a transverse field
$h_{f}\neq h_{i}$. We compare our results with the analytical solution
obtained from fermionization of the TFI model for a one-dimensional
chain with PBC. In the left panel of Fig. \ref{fig:Time-Accuracy}
the exact results for the time-dependent transverse spin polarization
are compared to NQS with $\alpha=4$. In the AFH model, we study instead
quantum quenches in the longitudinal coupling $J_{z}$ and monitor
the time evolution of the nearest-neighbors correlations. Our results
for the time evolution (and with $\alpha=4$ ) are compared with the
numerically-exact MPS dynamics \cite{white2004realtime,vidal2004efficient,daley2004timedependent}
for a system with open boundaries (see Fig. \ref{fig:Time-Accuracy},
right panel).

The high accuracy obtained also for the unitary dynamics further confirms
that neural network-based approaches can be fruitfully used to solve
the quantum many-body problem not only for ground-state properties
but also to model the evolution induced by a complex set of excited
quantum states. It is all in all remarkable that a purely stochastic
approach can solve with arbitrary degree of accuracy a class of problems
which have been traditionally inaccessible to QMC methods for the
past $50$ years. The flexibility of the NQS representation indeed
allows for an effective solution of the infamous phase problem plaguing
the totality of existing exact stochastic schemes based on Feynman's
path integrals.

\paragraph*{Outlook --- }

Variational quantum states based on artificial neural networks can
be used to efficiently capture the complexity of entangled many-body
systems both in one a two dimensions. Despite the simplicity of the
restricted Boltzmann machines used here, very accurate results for
both ground-state and dynamical properties of prototypical spin models
can be readily obtained. Potentially many novel research lines can
be envisaged in the near future. For example, the inclusion of the
most recent advances in machine learning, like deep network architectures,
might be further beneficial to increase the expressive power of the
NQS. Furthermore, the extension of our approach to treat quantum systems
other than interacting spins is, in principle, straightforward. In
this respect, applications to answer the most challenging questions
concerning interacting fermions in two-dimensions can already be anticipated.
Finally, at variance with Tensor Network States, the NQS feature intrinsically
non-local correlations which can lead to substantially more compact
representations of many-body quantum states. A formal analysis of
the NQS entanglement properties might therefore bring about substantially
new concepts in quantum information theory.
\begin{acknowledgments}
We acknowledge discussions with F. Becca, J.F. Carrasquilla, M. Dolfi,
J. Osorio, D. Patan�, and S. Sorella. The time-dependent MPS results
have been obtained with the open-source ALPS implementation \cite{bauer2011thealps,dolfi2014matrixproduct}.
This work was supported by the European Research Council through ERC
Advanced Grant SIMCOFE by the Swiss National Science Foundation through
NCCR QSIT, and by Microsoft Research. This paper is based upon work
supported in part by ODNI, IARPA via MIT Lincoln Laboratory Air Force
Contract No. FA8721-05-C-0002. The views and conclusions contained
herein are those of the authors and should not be interpreted as necessarily
representing the official policies or endorsements, either expressed
or implied, of ODNI, IARPA, or the U.S. Government. The U.S. Government
is authorized to reproduce and distribute reprints for Governmental
purpose not-withstanding any copyright annotation thereon.
\end{acknowledgments}

 \pagebreak{}
 \clearpage
\appendix

\section{Stochastic Optimization For The Ground State}

\begin{figure*}
\includegraphics[clip,width=1.7\columnwidth]{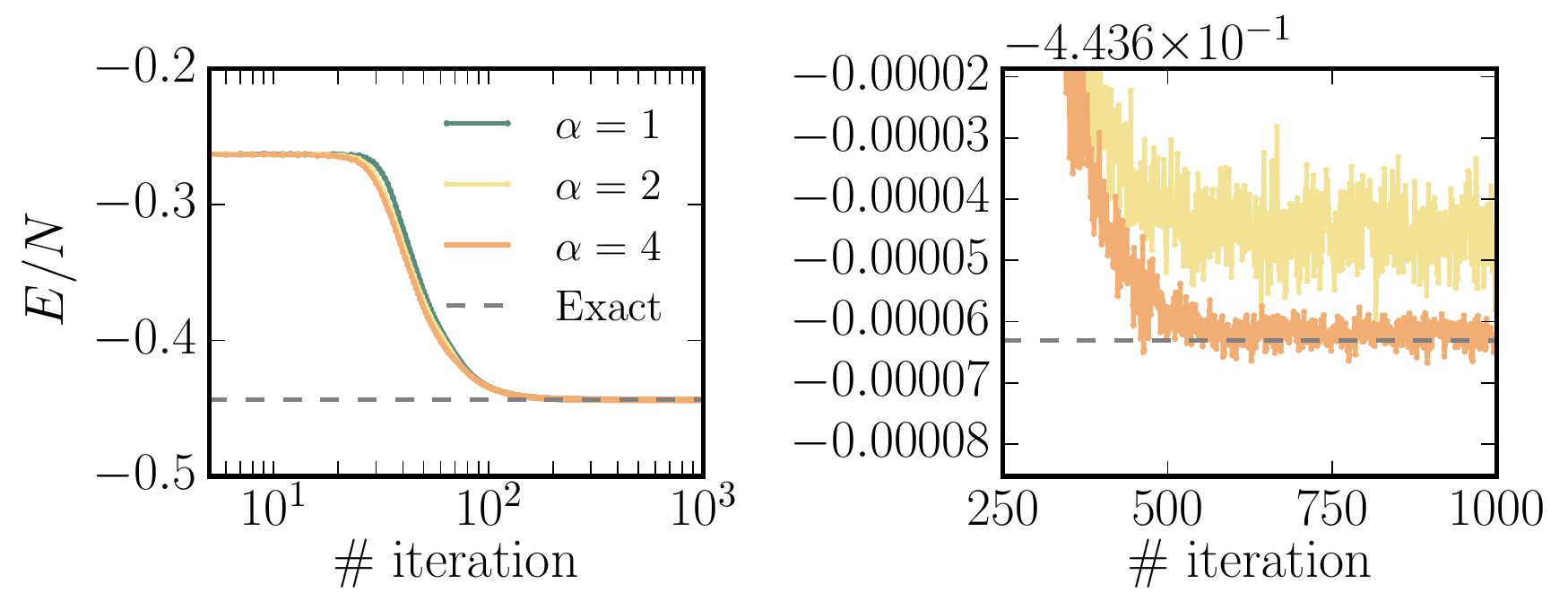}

\caption{\label{fig:Sr-Convergence}\textbf{Convergence properties of the stochastic
optimization.} Variational energy for the 1D Heisenberg model as a
function of the Stochastic Reconfiguration iterates, and for different
values of the hidden units density $\alpha$. The system has PBC over
a chain of $N=40$ spins. The energy converges smoothly to the exact
energy (dashed horizontal line) upon increasing $\alpha$. In the
Left panel we show a complete view of the optimization procedure and
on the Right panel a zoom in the neighborhood of the exact energy.}
\end{figure*}

In the first part of our Paper we have considered the goal of finding
the best representation of the ground state of a given quantum Hamiltonian
$\mathcal{H}$. The expectation value over our variational states
$E(\mathcal{W})=\langle\Psi_{M}|\mathcal{H}|\Psi_{M}\rangle/\langle\Psi_{M}|\Psi_{M}\rangle$
is a functional of the network weights $\mathcal{W}$. In order to
obtain an optimal solution for which $\nabla E(\mathcal{W}^{\star})=0$,
several optimization approaches can be used. Here, we have found convenient
to adopt the Stochastic Reconfiguration (SR) method of Sorella et
al. \cite{sorella2007weakbinding}, which can be interpreted as an
effective imaginary-time evolution in the variational subspace. Introducing
the variational derivatives with respect to the $k$-th network parameter,
\begin{eqnarray}
\mathcal{O}_{k}(\mathcal{S}) & = & \frac{1}{\Psi_{M}(\mathcal{S})}\partial_{\mathcal{W}_{k}}\Psi_{M}(\mathcal{S}),\label{eq:ovar}
\end{eqnarray}
as well as the so-called local energy
\begin{eqnarray}
E_{\textrm{loc}}(\mathcal{S}) & = & \frac{\langle\mathcal{S}|\mathcal{H}|\Psi_{M}\rangle}{\Psi_{M}(\mathcal{S})},\label{eq:localenergy}
\end{eqnarray}
the SR updates at the $p-$th iteration are of the form
\begin{equation}
\mathcal{W}(p+1)=\mathcal{W}(p)-\gamma S^{-1}(p)F(p),\label{eq:sreq}
\end{equation}
where we have introduced the (positive-definite) covariance matrix
\begin{eqnarray}
S_{kk^{\prime}}(p) & = & \langle\mathcal{O}_{k}^{\star}\mathcal{O}_{k^{\prime}}\rangle-\left\langle \mathcal{O}_{k}^{\star}\right\rangle \left\langle \mathcal{O}_{k^{\prime}}\right\rangle ,\label{eq:smatrix}
\end{eqnarray}
the \emph{forces }
\begin{eqnarray}
F_{k}(p) & = & \langle E_{\textrm{loc}}\mathcal{O}_{k}^{\star}\rangle-\langle E_{\textrm{loc}}\rangle\langle\mathcal{O}_{k}^{\star}\rangle,\label{eq:forces}
\end{eqnarray}
and a scaling parameter $\gamma(p)$. Since the covariance matrix
can be non-invertible, $S^{-1}$ denotes its Moore-Penrose pseudo-inverse.
Alternatively, an explicit regularization can be applied, of the form
$S_{k,k^{\prime}}^{\text{reg}}=S_{k,k^{\prime}}+\lambda(p)\delta_{k,k^{\prime}}S_{k,k}$
. In our work we have preferred the latter regularization, with a
decaying parameter $\lambda(p)=\max(\lambda_{0}b^{p},\lambda_{\text{min}})$
and typically take $\lambda_{0}=100$, $b=0.9$ and $\lambda_{\text{min}}=10^{-4}$.

Initially the network weights $\mathcal{W}$ are set to some small
random numbers and then optimized with the procedure outlined above.
In Fig. \ref{fig:Sr-Convergence} we show the typical behavior of
the optimization algorithm, which systematically approaches the exact
energy upon increasing the hidden units density $\alpha$.

\section{Time-Dependent Variational Monte Carlo}

In the second part of our Paper we have considered the problem of
solving the many-body Schr\"{o}dinger equation with a variational ansatz
of the NQS form. This task can be efficiently accomplished by means
of the Time-Dependent Variational Monte Carlo (t-VMC) method of Carleo
et al.

In particular, the \emph{residuals}
\begin{equation}
R(t;\dot{\mathcal{W}}(t))=\mathrm{dist}(\partial_{t}\Psi(\mathcal{W}(t)),-i\mathcal{H}\Psi)\label{eq:residuals}
\end{equation}
are a functional of the variational parameters derivatives, $\dot{\mathcal{W}}(t)$,
and can be interpreted as the quantum distance between the exactly-evolved
state and the variationally evolved one. Since in general we work
with unnormalized quantum states, the correct Hilbert-space distance
is given by the Fubini-Study metrics, given by
\begin{eqnarray}
\mathrm{dist_{\text{FS}}}(\Phi,\Phi^{\prime}) & = & \arccos\sqrt{\frac{\left\langle \Phi^{\prime}\right.\left|\Phi\right\rangle \left\langle \Phi\right.\left|\Phi^{\prime}\right\rangle }{\left\langle \Phi^{\prime}\right.\left|\Phi^{\prime}\right\rangle \left\langle \Phi\right.\left|\Phi\right\rangle }}.\label{eq:fubinistudy}
\end{eqnarray}
The explicit form of the residuals is then obtained considering $\Phi=\Psi+\delta\partial_{t}\Psi(\mathcal{W}(t))$
and $\Phi^{'}=\Psi-i\delta\mathcal{H}\Psi(\mathcal{W}(t))$. Taking
the lowest order in the time-step $\delta$ and explicitly minimizing
$\mathrm{dist_{\text{FS}}}(\Phi,\Phi^{\prime})^{2}$, yields the equations
of motion
\begin{eqnarray}
\dot{\mathcal{W}}(t) & = & -iS^{-1}(t)F(t),\label{eq:tvmceq}
\end{eqnarray}
where the correlation matrix and the forces are defined analogously
to the previous section. In this case the diagonal regularization,
in general, cannot be applied, and $S^{-1}(t)$ strictly denotes the
Moore-Penrose pseudo-inverse.

\begin{figure*}[t]
\includegraphics[clip,width=1.7\columnwidth]{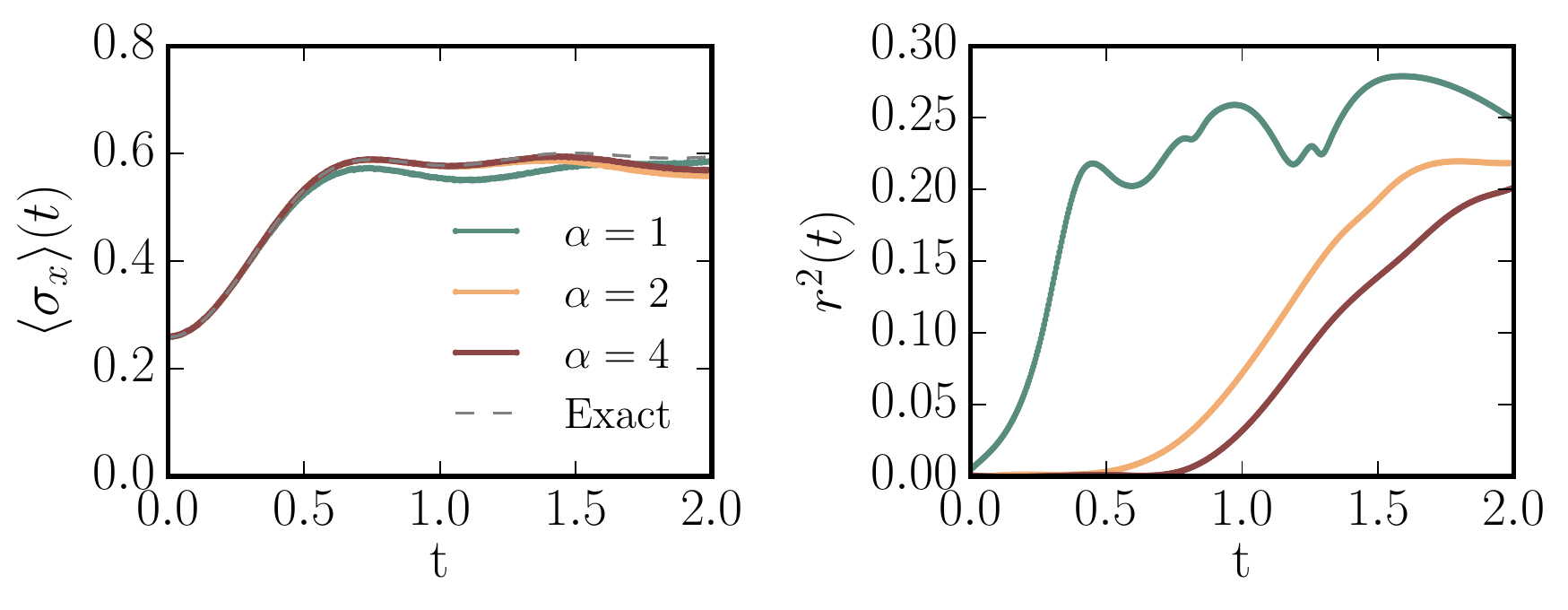}

\caption{\label{fig:Tvmc-Convergence}\textbf{Convergence properties of the
stochastic unitary evolution.} Time-dependent expectation value of
the transverse polarization along the $x$ direction in the TFI model,
for a quantum quench from $h_{i}=1/2$ to the critical interaction
$h_{f}=1$. t-VMC results are shown for different values of the hidden
units density $\alpha$. The system has periodic boundary conditions
over a chain of $N=40$ spins. (Left panel) The variational curves
for the expectation value of the transverse polarization converge
smoothly to the exact solution (dashed line) upon increasing $\alpha$.
(Right panel) The relative residual error $r^{2}(t)=R^{2}(t)/D_{0}^{2}(t)$,
where $D_{0}^{2}(t)=\textrm{dist}_{\textrm{FS}}(\Phi,\Phi-i\delta\mathcal{H})^{2}$
is shown for different values of the hidden unit density, and it is
systematically reduced increasing $\alpha$.}
\end{figure*}

The outlined procedure is globally stable as also already proven for
other wave functions in past works using the t-VMC approach. In Fig.
\ref{fig:Tvmc-Convergence} we show the typical behavior of the time-evolved
physical properties of interest, which systematically approach the
exact results when increasing $\alpha$.

\section{Efficient Stochastic Sampling}

We complete the supplementary information giving an explicit expression
for the variational derivatives previously introduced and of the overall
computational cost of the stochastic sampling. We start rewriting
the NQS in the form
\begin{eqnarray}
\Psi_{M}(\mathcal{S}) & = & e^{\sum_{i}a_{i}\sigma_{i}^{z}}\times\Pi_{j=1}^{M}2\cosh\theta_{j}(\mathcal{S}),\label{eq:PsimAppendix}
\end{eqnarray}
with the effective \emph{angles
\begin{equation}
\theta_{j}(\mathcal{S})=b_{j}+\sum_{i}W_{ij}\sigma_{i}^{z}.\label{eq:angles}
\end{equation}
}The derivatives then read
\begin{eqnarray}
\frac{1}{\Psi_{M}(\mathcal{S})}\partial_{a_{i}}\Psi_{M}(\mathcal{S}) & = & \sigma_{i}^{z},\label{eq:der1}\\
\frac{1}{\Psi_{M}(\mathcal{S})}\partial_{b_{j}}\Psi_{M}(\mathcal{S}) & = & \tanh\left[\theta_{j}(\mathcal{S})\right],\label{eq:der2}\\
\frac{1}{\Psi_{M}(\mathcal{S})}\partial_{W_{ij}}\Psi_{M}(\mathcal{S}) & = & \sigma_{i}^{z}\tanh\left[\theta_{j}(\mathcal{S})\right].\label{eq:der3}
\end{eqnarray}

In our stochastic procedure, we generate a Markov chain of many-body
configurations $\mathcal{S}^{(1)}\rightarrow\mathcal{S}^{(2)}\rightarrow\dots\mathcal{S}^{(P)}$
sampling the square modulus of the wave function $\left|\Psi_{M}(\mathcal{S})\right|^{2}$
for a given set of variational parameters. This task can be achieved
through a simple Metropolis-Hastings algorithm \cite{metropolis1953equation},
in which at each step of the Markov chain a random spin $s$ is flipped
and the new configuration accepted according to the probability
\begin{eqnarray}
A(\mathcal{S}^{(k)}\rightarrow\mathcal{S}^{(k+1)}) & = & \textrm{min}\left(1,\left|\frac{\Psi_{M}(\mathcal{S}^{(k+1)})}{\Psi_{M}(\mathcal{S}^{(k)})}\right|^{2}\right).\label{eq:acceptance}
\end{eqnarray}
In order to efficiently compute these acceptances, as well as the
variational derivatives, it is useful to keep in memory look-up tables
for the effective angles $\theta_{j}(\mathcal{S}^{(k)})$ and update
them when a new configuration is accepted. These are updated according
to
\begin{eqnarray}
\theta_{j}(\mathcal{S}^{(k+1)}) & = & \theta_{j}(\mathcal{S}^{(k)})-2W_{kj}\sigma_{s}^{z},\label{eq:thetas}
\end{eqnarray}
when the spin $s$ has been flipped. The overall cost of a Monte Carlo
sweep (i.e. of $\mathcal{O}(N)$ single-spin flip moves) is therefore
$\mathcal{O}(N\times M)=\mathcal{O}(\alpha N^{2}).$ Notice that the
computation of the variational derivatives comes at the same computational
cost as well as the computation of the local energies after a Monte
Carlo sweep.

\section{Iterative Solver}

The most time-consuming part of both the SR optimization and of the
t-VMC method is the solution of the linear systems (\ref{eq:sreq}
and \ref{eq:tvmceq}) in the presence of a large number of variational
parameters $N_{\textrm{var}}$. Explicitly forming the correlation
matrix $S$, via stochastic sampling, has a dominant quadratic cost
in the number of variational parameters, $\mathcal{O}(N_{\textrm{var}}^{2}\times N_{\textrm{MC}})$,
where $N_{\textrm{MC}}$ denotes the number of Monte Carlo sweeps.
However, this cost can be significantly reduced by means of iterative
solvers which never form the covariance matrix explicitly. In particular,
we adopt the MINRES-QLP method of Choi and Saunders \cite{choi2014algorithm},
which implements a modified conjugate-gradient iteration based on
Lanczos tridiagonalization. This method iteratively computes the pseudo-inverse
$S^{-1}$ within numerical precision. The backbone of iterative solvers
is, in general, the application of the matrix to be inverted to a
given (test) vector. This can be efficiently implemented due to the
product structure of the covariance matrix, and determines a dominant
complexity of $\mathcal{O}(N_{\textrm{var}}\times N_{\textrm{MC}})$
operations for the sparse solver. For example, in the most challenging
case when translational symmetry is absent, we have $N_{\textrm{var}}=\alpha N^{2}$,
and the dominant computational cost for solving (\ref{eq:sreq} and
\ref{eq:tvmceq}) is in line with the complexity of the previously
described Monte Carlo sampling.

\section{Implementing Symmetries }

Very often, physical Hamiltonians exhibit intrinsic symmetries which
must be satisfied also by their ground- and dynamically-evolved quantum
states. These symmetries can be conveniently used to reduce the number
of variational parameters in the NQS.

Let us consider a symmetry group defined by a set of linear transformations
$T_{s}$, with $s=1,\dots S$, such that the spin configurations transform
according to $T_{s}\sigma^{z}=\tilde{\sigma}^{z}(s)$. We can enforce
the NQS representation to be invariant under the action of $T$ defining
\begin{multline}
\Psi_{\alpha}(\mathcal{S};\mathcal{W})=\sum_{\{h_{i,s}\}}\exp\left[\sum_{f}^{\alpha}a^{(s)}\sum_{s}^{S}\sum_{j}^{N}\tilde{\sigma_{j}^{z}}(s)+\right.\\
\left.+\sum_{f}^{\alpha}b^{(s)}\sum_{s}^{S}h_{f,s}+\sum_{f}^{\alpha}\sum_{s}^{S}h_{f,s}\sum_{j}^{N}W_{j}^{(f)}\tilde{\sigma_{j}^{z}}(s)\right],\label{eq:PsiInvariant}
\end{multline}
where the network weights have now a different dimension with respect
to the standard NQS. In particular, $a^{(f)}$ and $b^{(f)}$ are
vectors in the \emph{feature} space with $f=1,\dots\alpha_{s}$ and
the connectivity matrix $W_{j}^{(f)}$ contains $\alpha_{s}\times N$
elements. Notice that this expression corresponds effectively to a
standard NQS with $M=S\times\alpha_{s}$ hidden variables. Tracing
out explicitly the hidden variables, we obtain
\begin{multline}
\Psi_{\alpha}(\mathcal{S};\mathcal{W})=e^{\sum_{f,s,j}a^{(f)}\tilde{\sigma_{j}^{z}}(s)}\times\\
\times\Pi_{f}\Pi_{s}2\cosh\left[b^{(f)}+\sum_{j}^{N}W_{j}^{(f)}\tilde{\sigma_{j}^{z}}(s)\right].\label{eq:PsiInvariantSummed}
\end{multline}
In the specific case of site translation invariance, we have that
the symmetry group has an orbit of $S=N$ elements. For a given feature
$f$, the matrix $W_{j}^{(f)}$ can be seen as a filter acting on
the $N$ translated copies of a given spin configuration. In other
words, each feature has a pool of $N$ associated hidden variables
that act with the same filter on the symmetry-transformed images of
the spins.
\end{document}